\documentstyle[eqsecnum,prd,aps,epsf]{revtex}

\begin{document}

%%%%%%%%%%%%%%%%%%%%%%%%%%%%%%%%%%%%%%%%%%%%%%%%%%%%%%%%%%%%%%
%% O. FRONT MATTER                                          %%
%%%%%%%%%%%%%%%%%%%%%%%%%%%%%%%%%%%%%%%%%%%%%%%%%%%%%%%%%%%%%%

\begin{center}
{\large\bf{$R$-mode Instability of Slowly Rotating Non-isentropic 
Relativistic Stars
}}
~\\
~\\
Shijun Yoshida 
and 
Toshifumi Futamase \\
~\\
{\em Astronomical Institute, Graduate School of Science, 
Tohoku University, Sendai 980-8578, Japan 
}
\end{center}

\begin{abstract}

We investigate properties of $r$-mode instability 
in slowly rotating relativistic polytropes. Inside the star 
slow rotation and low frequency formalism that was mainly 
developed by Kojima is employed to study axial oscillations 
restored by Coriolis force. At the stellar surface, in order 
to take account of gravitational radiation reaction effect, 
we use a near-zone boundary condition instead of the usually 
imposed boundary condition for asymptotically flat 
spacetime. Due to the boundary condition, complex frequencies 
whose imaginary part represents secular instability are 
obtained for discrete $r$-mode 
oscillations in some polytropic models. It is found that such discrete 
$r$-mode solutions can be obtained only for some restricted polytropic 
models. Basic properties of the solutions are similar to those obtained by 
imposing the boundary condition for asymptotically flat spacetime. Our 
results suggest that existence of a continuous part of spectrum cannot be 
avoided even when its frequency becomes complex due to the emission of
gravitational radiation. 
 
\end{abstract}


%%%%%%%%%%%%%%%%%%%%%%%%%%%%%%%%%%%%%%%%%%%%%%%%%%%%%%%%%%%%%%
%% I. INTRODUCTION                                          %%
%%%%%%%%%%%%%%%%%%%%%%%%%%%%%%%%%%%%%%%%%%%%%%%%%%%%%%%%%%%%%%

\section{Introduction}

Andersson \cite{nils97} and Friedman and Morsink \cite{jfs97} discovered 
that all  
$r$-modes, which are quasi-toroidal modes mainly restored by 
Coriolis force, in all rotating stars become unstable due to the 
gravitational radiation reaction if other dissipative processes are 
not considered. This instability is clearly understood by the so called 
CFS mechanism \cite{ch70,fs78,fr78}. 
As shown by Lindblom, Owen, and Morsink \cite{lom98} first, this instability 
still strongly affects on stability of typical neutron star models even 
if viscous dissipation of neutron star matter, which tends to stabilize 
the CFS instability, is taken into account. Since then a lot of studies 
on oscillation modes restored by Coriolis force in rotating stars 
have been done to prove their 
possible importance in astrophysics (for recent review, see, e.g., 
Refs \cite{fl99,ak00,li01,fl01}).

Influence of the $r$-mode on stability of rotating neutron stars is one 
of the most important and interesting phenomenon in astrophysics.  
In oscillations of neutron stars, relativistic effect 
must be important because such stars are sufficiently compact. But most 
studies have been done within the framework of Newtonian gravity so far, 
although our understandings of $r$-modes have been improved by those 
investigations. As for $r$-modes studied within the framework 
of general relativity, Kojima \cite{k98} derived master equations for $r$-mode 
oscillation in the lowest order slow rotation approximation, and then 
he found possible existence of a continuous part of spectrum in his equations.  
Beyer and Kokkotas \cite{bk99} generally verified the existence of a continuous 
part of spectrum in Kojima's equation. Kojima's formalism was developed 
to include higher order rotational effects by Kojima and Hosonuma 
\cite{kh99,kh00}. Lockitch, Andersson, and Friedman \cite{laf01} obtained the 
discrete $r$-mode 
solutions in uniform density stars as well as a continuous 
part of spectrum by solving Kojima's equation. Recently Yoshida \cite{y01} and 
Ruoff and Kokkotas \cite{rk01} discussed that such discrete $r$-mode solutions 
are not simply allowed to appear in compressible stellar models. Their 
results showed that for typical neutron star models, Kojima's equations 
do not have such a discrete $r$-mode solution.

These recent developments of understanding of relativistic $r$-modes have 
shown that basic properties of $r$-mode oscillations in relativistic stars 
are significantly different from those in Newtonian stars. As for 
non-isentropic stars, most previous studies have shown the existence of a 
continuous part of spectrum. This is a great contrast with  
Newtonian case. For Newtonian cases, there are discrete mode solutions and 
no continuous parts of spectrum for $r$-modes in all uniformly rotating 
stars as long as their rotation velocity is small enough.  
It is not however likely to occur such drastic change in the behavior of the 
solutions due to the inclusion of even a little relativistic effect. Therefore, 
most authors have considered that a continuous part of spectrum does not 
appear if some effects that were omitted in previous studies are taken 
into account. One of such effects is a dissipation effect due to gravitational 
radiation reaction. In most studies on $r$-modes, a slow motion approximation 
has been employed in their analysis because of slow rotation approximation. 
A slow 
motion approximation changes a wave type equation into a Laplace type equation. 
Thus, Kojima's equations do not have solutions with wave character, and 
then the frequency 
is real number if asymptotically flat spacetime is assumed. In other words, 
influence of gravitational radiation on relativistic $r$-modes has not been 
taken into account so far. In this case, 
Kojima's equation becomes that of singular eigenvalue problem for some 
frequency range, and hence has a continuous part of spectrum. 
As suggested by Lockitch et al. \cite{laf01} (see also, Ref. \cite{bk99}), 
however, Kojima's equation may become that of regular eigenvalue 
problem if frequency has non-zero imaginary part. Therefore, It is hoped that 
the existence 
of a continuous part of spectrum might be avoided if the effect of the 
emission of the gravitational radiation is considered.

In this paper, accordingly, we will attempt to include the lowest order 
effect of gravitational radiation reaction into Kojima's formalism.  
Because Kojima's equation is not wave type equation, as mentioned before, 
we can not obtain information about gravitational wave from the equation 
at all. In order to include the gravitational radiation reaction effect 
into Kojima's equation, we will employ a near-zone boundary condition 
instead of usually used boundary condition for asymptotically flat spacetime. 
This boundary condition was introduced by Thorne \cite{tc69} to include the 
effect of gravitational 
radiation reaction on polar pulsations in a Newtonian star.  
We start, in \S 2, with the description 
of our method of solution. A near-zone boundary condition is introduced 
there in order to take account of the gravitational radiation reaction. 
In \S 3, we show the properties of the $r$-mode 
solutions obtained by imposing the boundary condition.  
\S 4 is devoted for discussions and conclusions. Throughout this paper we 
will use units in which $c=G=1$, where $c$ and $G$ denote the velocity of 
light and the gravitational constant, respectively.

%%%%%%%%%%%%%%%%%%%%%%%%%%%%%%%%%%%%%%%%%%%%%%%%%%%%%%%%%%%%%%
%% II. Method of solutions                                  %%
%%%%%%%%%%%%%%%%%%%%%%%%%%%%%%%%%%%%%%%%%%%%%%%%%%%%%%%%%%%%%%

\section{Method of solutions}

We consider slowly rotating relativistic stars with a uniform angular 
velocity $\Omega$, where we take account of the first order rotational 
effect in $\Omega$. The geometry around the equilibrium stars can be 
described by the following line element (see, e.g. Ref. \cite{th71}): 
\begin{eqnarray}
ds^2 = - e^{2 \nu(r)} dt^2 + e^{2 \lambda(r)} dr^2 + r^2 d\theta^2 + 
           r^2 \sin^2 \theta d\varphi^2 - 
           2 \omega(r) \, r^2 \sin^2 \theta dt d\varphi  \, . 
\label{metric}
\end{eqnarray}
Throughout this paper, the polytropic equation of state is assumed:   
\begin{equation}
p = K\, \rho^{1+\frac{1}{N}} \, ,
\end{equation}
where $p$ and $\rho$ denote the pressure, mass-energy density, respectively.
Here $N$ and $K$ are constants.

We consider oscillation modes in rotating relativistic stars such that 
the eigenfunctions are stationary and are composed of only one axial parity 
component in the limit $\Omega\rightarrow0$. Subclass of those modes should 
be a relativistic counterpart of $r$-modes, which are able to oscillate in 
all slowly rotating Newtonian fluid stars. According to Lockitch et al. 
\cite{laf01}, such modes are allowed to exist only if the star has 
non-isentropic 
structures. Therefore, we assume stars to be non-isentropic, although the 
effects due to deviation from isentropic structure on oscillation modes 
do not appear in the first order in $\Omega$. According to the formalism 
by Lockitch et al. \cite{laf01} (see, also, Ref. \cite{k98}), 
let us write down the pulsation equation for 
relativistic $r$-modes with accuracy up to the first order in $\Omega$. 
The metric perturbation $\delta g_{\alpha\beta}$ and the Eulerian changes  
of the fluid velocity $\delta u^\alpha$ that do not vanish in the limit 
$\Omega\rightarrow0$ are given as 
\begin{equation}
(\delta g_{t\theta},\delta g_{t\varphi})=i h_{0,l}(r) \, 
\left(-\frac{\partial_\varphi Y_{lm}(\theta,\varphi)}{\sin\theta}\, ,
\sin\theta\partial_\theta Y_{lm}(\theta,\varphi)\right)\, e^{i \sigma t} \, , 
\label{ex-g}
\end{equation}
\begin{equation}
(\delta u^\theta,\delta u^\varphi)=\frac{i U_{l}(r)}{r^2} \, 
\left(-\frac{\partial_\varphi Y_{lm}(\theta,\varphi)}{\sin\theta}\, ,
\frac{\partial_\theta Y_{lm}(\theta,\varphi)}{\sin\theta}\right)\, 
e^{i \sigma t} \, ,
\label{ex-v}
\end{equation}
where $Y_{lm}(\theta,\varphi)$ are the usual spherical harmonic functions, and 
$\sigma$ denotes oscillation frequency measured in the inertial frame at the 
spatial infinity. All other perturbed quantities become higher order in  
$\Omega$. Note that this form of eigenfunctions is the same as that for 
zero-frequency modes in a spherical non-isentropic star, because $r$-modes 
become zero-frequency ones in the limit $\Omega\rightarrow0$ \cite{tc67}. 
The metric perturbation $h_{0,l}$ obeys a second order 
ordinary differential equation,   
\begin{eqnarray}
D_{lm}(r;\bar{\sigma}) \, 
\left[e^{\nu-\lambda}\frac{d}{dr}
\left(e^{-\nu-\lambda}\frac{dh_{0,l}}{dr}\right)-\left(\frac{l(l+1)}{r^2}+
\frac{ -2 + 2 e^{-2 \lambda}}{r^2}+8 \pi (p+\rho)\right)h_{0,l}\right] 
+ 16 \pi (p+\rho) h_{0,l}=0 \, ,
\label{b-eq}
\end{eqnarray}
where 
\begin{equation}
D_{lm}(r;\bar{\sigma})\equiv 1-\frac{2 m \bar{\omega}}{l(l+1)\bar{\sigma}} \,
,
\label{coe1}
\end{equation}
Here, we have introduced the effective rotation angular velocity of fluid,  
\begin{equation}
\bar{\omega} = \Omega - \omega \, , 
\end{equation}
and the corotating oscillation frequency, 
\begin{equation}
\bar{\sigma} = \sigma + m \Omega \, .
\end{equation}
The velocity perturbation of fluids $U_l$ is determined from the function 
$h_{0,l}$ through the following algebraic relation, 
\begin{equation}
\left[1-\frac{2 m \bar{\omega}}{l(l+1)\bar{\sigma}}\right]\, U_{l}+h_{0,l}=0 
\, . 
\label{b-eq2}
\end{equation}
Equations (\ref{b-eq}) and (\ref{b-eq2}) are our basic equations, which 
were derived by Kojima \cite{k98} first. 
Note that the equations 
(\ref{b-eq}) and (\ref{b-eq2}) lose their meaning in the $l=0$ case, because 
there are no axial modes with $l=0$.

Because equations (\ref{b-eq}) are second order ordinary differential 
equations, two boundary conditions are required to determine solutions 
uniquely. From regularity of physical quantities at $r=0$ the function 
$h_{0,l}$ must vanish at the center of a star. This condition is explicitly 
given as 
\begin{eqnarray}
r \, \frac{d h_{0,l}}{dr} - (l+1)\, h_{0,l} = 0  \ \ \ 
{\rm as} \ r\rightarrow 0 \, . 
\label{bc1}
\end{eqnarray}
Outside the star equations (\ref{b-eq}) have general solutions as follows:
\begin{eqnarray}
h_{0,l} (r) =  A \sum_{k=0}^\infty a_k r^{-l-k} + 
               B \sum_{k=0}^\infty b_k r^{l+1+k} \, ,
\end{eqnarray}
where $A$ and $B$ are arbitrary constants. Here $a_k$ and $b_k$ are 
constants determined from recurrence relations although explicit 
expressions are omitted here. Note that $a_k$ and $b_k$ do not depend on 
frequency $\bar{\sigma}$. In most previous studies the condition 
$B=0$ have been chosen as a boundary  
condition at the spatial infinity because spacetime must be regular 
everywhere. In this paper we call this condition ``proper boundary 
condition''. This condition means the spacetime is asymptotically flat.   
Therefore, solutions satisfying the condition $B=0$ can not 
describe any gravitational radiation emission from a star at all. 
In order to include effect of gravitational radiation emission in 
the solutions, we must not require to demand the zero for $B$. 
This was first shown and used in the study for post-Newtonian 
approximation by Thorne \cite{tc69}.

Next, let us consider the boundary conditions for quasi-normal mode 
solutions, according to similar consideration to that by Thorne \cite{tc69}  
(see, also, Ref. \cite{lmi97}). 
In the derivation of our basic equations slow motion approximation is 
necessarily used as well as slow rotation one because the frequency of 
oscillation restored by Coriolis force is the same order of the stellar 
rotation frequency. This slow motion approximation must nicely work near 
the star as long as low frequency oscillations are considered. 
Therefore, higher order time derivative of perturbed quantities are 
neglected in governing equations (\ref{b-eq}).  If an oscillating 
star emits 
gravitational radiation, however, some of such omitted terms must become 
important in the radiation zone because a term such as $\sigma r$ becomes 
dominant among all terms in the governing equations. Besides, 
rotational effects due to the stellar rotation fall off faster than $1/r^2$ 
as $r\rightarrow \infty$. Thus, such slow motion approximation is not good 
far from the star, if gravitational wave are radiated from the stars. 
Regge-Wheeler equations with correction terms due to the stellar rotation, 
in fact, govern the axial perturbations sufficiently far from the star even 
when low frequency oscillations like $r$-modes are induced. Since 
$\sigma r \gg 1$ in radiation zone, Regge-Wheeler equations can be 
approximately written as 
\begin{eqnarray}
r^2 \, \frac{d^2 X_l(r)}{dr^2} + \sigma^2 r^2 X_l(r) -l(l+1) X_l(r) 
+ O\left(\frac{M}{r}\right) = 0 \, ,
\label{RW-eq}
\end{eqnarray}
where $X_l(r)$ are Regge-Wheeler's functions. Here $M$ denotes the 
gravitational mass of the star. The metric functions 
$h_{0,l}$ are determined from the functions $X_l$ by the equation: 
\begin{eqnarray}
h_{0,l} = \frac{d (r X_l(r))}{dr} + O\left(\frac{M}{r}\right) \, , 
\label{rel-hX}
\end{eqnarray}
(see, e.g. Refs \cite{rw57,nok87}). 
Here we have considered the limiting case when $M/r \ll 1$ for simplicity. 
The general solutions to equations (\ref{RW-eq}) can be given analytically: 
\begin{eqnarray}
X_l (\sigma r)= \sigma r \, (C \, j_l (\sigma r) 
                              + D \, n_l (\sigma r) ) \, , 
\label{ana-sol}
\end{eqnarray}
where $j_l$ and $n_l$ are spherical Bessel functions and $C$ and $D$ are 
arbitrary constants. The asymptotic forms in the radiation zone, that is, 
when $\sigma r \gg 1$, are given as 
\begin{eqnarray}
X_l (\sigma r)\sim C \, \cos\left[\sigma r - \frac{1}{2}\,(l+1)\pi\right]
    + D \, \sin\left[\sigma r - \frac{1}{2}\,(l+1)\pi\right] \, . 
\label{farzone}
\end{eqnarray}
Now we are interested in quasi-normal modes of a star. Thus, the no 
incoming radiation conditions are required for metric perturbations. 
From equation (\ref{farzone}) it is found that this condition becomes  
the relation $D=-i\,C$. For this choice of constants the asymptotic 
solutions reduce to the form as: 
\begin{eqnarray}
X_l (\sigma r) \, e^{i \sigma t}\sim C \, \exp 
 \left[i \sigma (t-r) + \frac{i}{2}\,(l+1)\pi \right] \, , 
 \ \ \ \ \ {\rm as} \ \ \sigma r \rightarrow \infty \, .  
\end{eqnarray}
Since the frequencies of the oscillations restored by the 
Coriolis force are proportional to the rotational frequency $\Omega$,  
$\sigma^2 R^2 \approx \Omega^2 R^2 = \epsilon^2 M/R$, where $\epsilon$ 
denotes a small parameter for stellar rotation defined as 
$\epsilon=\Omega/(M/R^3)^{1/2}$. Here, $R$ is the radius of the star. Thus, 
the surface of the star is approximately in the near zone, that is, 
$\sigma r \ll 1$, if the stellar rotation is sufficiently slow or the 
stellar gravity is sufficiently weak. In this approximation, solutions 
(\ref{ana-sol}) can be written as 
\begin{eqnarray}
X_l (\sigma r) \sim -i\,C \,\frac{(2l-1)!!}{(\sigma r)^{l}} 
 \left[ 1 + i \frac{(\sigma r)^{2 l+1}}{(2l-1)!!(2l+1)!!} \right] \, , 
\label{nearzone}
\end{eqnarray}
where the constraint $D=-i\,C$ for the no incoming radiation has been 
used. The approximate solutions above may be valid near the stellar surface 
because $M/r<1$ is 
well satisfied at the stellar surface and outside the star for typical
 neutron star models. In the near zone, thus, the expressions for metric 
perturbations $h_{0,l}$ outside the star can be given by equations 
(\ref{rel-hX}) and (\ref{nearzone}) as follows: 
\begin{eqnarray}
h_{0,l} \sim C'\,\frac{1}{r^l}\,
\left[1+i\frac{(l+2)(\sigma r)^{2 l+1}}{(l-1)(2l+1)[(2l-1)!!]^2}\right] 
\, ,
\label{outer-bc}
\end{eqnarray}
where $C'$ is an arbitrary constant. As the outer boundary condition we 
require this solution to connect smoothly to interior solution 
obtained by solving equation (\ref{b-eq}) at the stellar surface. Thus, 
the boundary condition is explicitly given as 
\begin{eqnarray}
&&\left[1+i\frac{(l+2)(\sigma R)^{2 l+1}}{(l-1)(2l+1)[(2l-1)!!]^2}\right] 
\, r\,\frac{d h_{0,l}}{dr}(R-x) \nonumber \\
&&+\left[l-i
\frac{(l+1)(l+2)(\sigma R)^{2 l+1}}{(l-1)(2l+1)[(2l-1)!!]^2}\right]
 \,h_{0,l}(R-x) = 0\, , \ \ \ {\rm as} \ x\rightarrow 0 \, ,
\label{nz-bc}
\end{eqnarray}
where we need actual values of rotation frequency $\Omega$ to 
obtain eigensolutions. In this paper we assume the value of $\Omega$ as 
$\Omega=(\pi \bar{\rho})^{1/2}$, where $\bar{\rho}$ is the average density 
defined by $\bar{\rho}=M/(4\pi R^3/3)$. This $\Omega$ is an approximate 
value for 
maximum rotation frequency to be possible to settle down uniformly 
rotating stars in hydrostatic equilibrium. In this paper, we call the 
condition (\ref{nz-bc}) ``near zone boundary condition''. This method is a 
crude version of matched asymptotic expansions. If we consider a 
non-rotating limit, that is, $\sigma=0$ limit, the boundary condition 
(\ref{nz-bc}) becomes approximation of proper boundary conditions in 
which only the lowest order terms in $M/R$ are included. As shown by 
Lindblom et al. \cite{lmi97}, the boundary condition similar to that by using 
asymptotic solution (\ref{outer-bc}) can give a good approximate value of 
eigenfrequency even for $f$-mode oscillation although $f$-mode is not a low 
frequency oscillation mode.

%%%%%%%%%%%%%%%%%%%%%%%%%%%%%%%%%%%%%%%%%%%%%%%%%%%%%%%%%%%%%%
%% III. Numerical results                                   %%
%%%%%%%%%%%%%%%%%%%%%%%%%%%%%%%%%%%%%%%%%%%%%%%%%%%%%%%%%%%%%%

\section{Numerical results}

As shown in the previous studies \cite{k98,bk99,laf01},  
we should distinguish two cases in treating 
equation (\ref{b-eq}) when the proper boundary condition is imposed. 
One case is regular eigenvalue problem and the other singular one. 
For regular eigenvalue problem equation (\ref{b-eq}) may have 
discrete mode frequency in a range, 
\begin{equation}
\frac{2 m \bar{\omega}(R)}{l(l+1)} < \bar{\sigma}  \le
\frac{2 m \bar{\omega}(\infty)}{l(l+1)} = 
\frac{2 m \Omega}{l(l+1)} \, .  
\label{dis-sp}
\end{equation}
On the other hand, equation (\ref{b-eq}) becomes singular 
eigenvalue problem if $\bar{\sigma}$ is in a region, 
\begin{equation}
\frac{2 m \bar{\omega}(0)}{l(l+1)} \le \bar{\sigma} \le 
\frac{2 m \bar{\omega}(R)}{l(l+1)} \, . 
\label{con-sp}
\end{equation}
Notice that the range (\ref{con-sp}) is a continuous part of spectrum 
of equation (\ref{b-eq}). As pointed out by Lockitch et al. \cite{laf01}
 (see also, Ref. \cite{bk99})
equation (\ref{b-eq}) becomes that of regular eigenvalue problem when the 
corresponding frequency has non-zero imaginary part. Thus frequency ranges  
above may not have clear mathematical meanings when frequency has non-zero 
imaginary part. According to frequency ranges above, however, we will divide 
eigensolutions into three classes: The first and the second 
class solutions are characterized by their real part of frequency 
satisfying inequality (\ref{dis-sp}) and (\ref{con-sp}), respectively. 
The third class composes of a compensative set of the first and the second 
class.

First of all, we concentrate our attention to $r$-mode solutions with 
frequency whose real part is in a range (\ref{dis-sp}). We compute 
frequencies of mode solutions for several polytropic stellar models. 
In the present study, only the fundamental $r$-modes, whose eigenfunction 
$U_m$ has no node in radial direction except at stellar center, are 
obtained. This is similar to that in studies for the proper boundary 
condition case \cite{y01,rk01}. In Figures 1 
and 2, real and imaginary parts of scaled eigenfrequencies 
$\kappa \equiv \bar{\sigma}/\Omega$ of $r$-modes are, respectively, given 
as functions of $M/R$. Eigenfrequencies for stars 
with four different polytropic indices, $N=0$, $0.5$, $0.75$, and $1$, 
are shown in panels in both figures, respectively. Only frequency curves 
for the modes with $l=m=2$ are depicted along a relativistic factor $M/R$ 
because they are considered to be the most important modes for $r$-mode 
instability.

Real parts of frequency illustrated in Figure 1 are in good agreement with 
Figure 1 of Ref. \cite{y01}, in which frequency are obtained by imposing 
the proper boundary condition for asymptotically flat spacetime. The 
relative differences are less than $0.1 \%$. This shows that our approximation 
works nicely and higher order effect of $M/r$ on the outer boundary condition 
is not so important for the determination of real parts of frequency. 
We also find that frequency curves in Figure 1 are terminated at some value 
of $M/R$ beyond which equilibrium states can still exist. Here, the 
maximum values of $M/R$ for polytropic equilibrium stars having $N=0$, 
$0.5$, $0.75$, and $1.0$ are given by $4/9$, $0.385$, $0.349$, and $0.312$, 
respectively. It is also found that length of frequency curves tend to be 
short as a polytrope index $N$ increases. This feature is similar to that 
for the case where the proper boundary condition is used. And those terminal 
points of frequency curves appear at almost the same values of relativistic 
factors as those for the proper boundary condition case (see Ref. \cite{y01}).  
Beyond the value of the relativistic factors corresponding to those terminal 
points, we can obtain a lot of eigensolutions with a singular eigenfunction 
but not with a regular one. Furthermore, the real part of the corresponding 
frequency belongs to a range (\ref{con-sp}) but not a range (\ref{dis-sp}).

From Figure 2, we can see that $r$-modes obtained in this study are all 
unstable. It is also found that curves for imaginary parts of $\kappa$ 
have one relative minimum near the terminal point of frequency curves. 
Those minimum values are given as 
${\rm Im}(\kappa(M/R=0.425))=-2.9\times 10^{-3}$, 
${\rm Im}(\kappa(M/R=0.344))=-6.4\times 10^{-4}$,
${\rm Im}(\kappa(M/R=0.258))=-1.3\times 10^{-4}$, and 
${\rm Im}(\kappa(M/R=0.095))=-2.2\times 10^{-6}$, for stars having $N=0$, 
$0.5$, $0.75$, and $1.0$, respectively. 
Values of ${\rm Im}(\kappa)$ also approach zero as relativistic factor $M/R$ 
is getting closer to a value corresponding to that for the terminal point of 
frequency curves. Those behaviors of ${\rm Im}(\kappa)$ can be understood 
from the distribution of eigenfunctions $U_l(r)$ because ${\rm Im}(\kappa)$ 
is approximately proportional to the square of the current multipole moment. 
In Figure 3 and 4 distributions of the eigenfunction $U_m$ are shown for 
$N=0.5$ polytropic models having $M/R=0.1$ and $M/R=0.37$, respectively. 
As seen from these figures the motion of perturbed fluid elements is 
strongly confined near the stellar surface when the mode frequency is 
getting closer to that of the terminal frequency, which satisfies 
$\bar{\sigma} \approx 2 m \bar{\omega}(R)/(l(l+1))$. We can easily 
understand this behavior from equation (\ref{b-eq2}). Consequently 
the values of current multipole moments of such modes may become small 
when the value of relativistic factor increases. On the other hand, 
the efficiency of the gravitational radiation emission becomes good as 
increase of the value of the relativistic factor. Due to both effects above  
a relative minimum of imaginary part of frequency may appear. We should 
notice that in $N=0$ case, a value of Im$(\kappa)$ do not approach  
zero even when $M/R \sim 4/9$, as we can see from Figure 2. The reason  
is that the eigenfunctions $U_m$ does not have so strong peak at the 
stellar surface even when $M/R \sim 4/9$ because a relation 
$\bar{\sigma} \approx 2 m \bar{\omega}(R)/(l(l+1))$ is not satisfied in 
this case.

Next let us estimate the instability timescale for the gravitational 
radiation driven instability of $r$-mode solutions with frequency 
in a range (\ref{dis-sp}). Here a typical neutron star model whose 
mass and radius are respectively $1.4 M_{\odot}$ and $12.57$ km 
is considered for $N=0$, $0.5$, and $0.75$ polytropic models. When 
the star rotates with the angular frequency 
$\Omega = (3/4 \times G M/R^3)^{1/2} =  8377 \ s^{-1}$, growing 
timescales  $\tau_j$ of the $r$-mode instability are given as 
$\tau_j = 1.29 \ s$, $2.04 \ s$, and $2.97 \ s$ for  $N=0$, $0.5$, 
and $0.75$ models, respectively. Note that for $N=1$ case the growing 
timescale cannot be estimated because we can not find discrete $r$-mode 
solution for that model. These timescale are similar to those obtained 
from Newtonian estimate. (see, e.g. Ref. \cite{lom98})

When we concentrate our attention to solutions whose frequency has real 
part in a frequency range (\ref{con-sp}), we obtain a large number of 
solutions whose eigenfunction has singular behavior in its real part at 
a radius determined by a solution of $D_{lm}(r;{\rm Re}(\bar{\sigma}))=0$. 
Besides, those solutions have severe truncation error due to a finite 
difference method. Similar behavior of solutions appears in oscillations of 
differentially rotating disks. (see, for example, Ref. \cite{sv83}) 
As discussed by Schutz and Verdaguer \cite{sv83}, this is considered a sign of 
the existence of continuous parts of spectrum, although the existence has 
to be proved by other mathematical techniques because exact continuous 
spectrum is never obtained from a simple numerical analysis. In the present 
case, we are sure that the appearance of a continuous spectrum is plausible 
because the imaginary parts of frequency are too small to change drastically 
the character of the solutions derived from the proper boundary condition. 
Thus, our numerical results suggest the existence of a continuous part of 
spectrum in Kojima's equation even when their frequency becomes complex 
number. As for regular solutions with frequency in a range (\ref{con-sp}), 
we cannot obtain such a solution at all. 
Finally we consider the third class of solutions, whose real part of 
frequency is neither in a range (\ref{dis-sp}) nor (\ref{con-sp}). In this 
region of real part of frequency, we cannot obtain any solutions at all.

%%%%%%%%%%%%%%%%%%%%%%%%%%%%%%%%%%%%%%%%%%%%%%%%%%%%%%%%%%%%%%
%%       DISCUSSIONS AND CONCLUSIONS                        %%
%%%%%%%%%%%%%%%%%%%%%%%%%%%%%%%%%%%%%%%%%%%%%%%%%%%%%%%%%%%%%%

\section{Discussion and Conclusion}

In this paper, we have investigated the properties of $r$-mode instability 
in slowly rotating relativistic polytropes. Inside the star slow rotation 
and the low frequency formalism that was mainly developed by Kojima \cite{k98} 
and Lockitch et al. \cite{laf01} is employed to study axial oscillations 
restored 
by Coriolis force. At the stellar surface, in order to take account of 
gravitational radiation reaction effect, we use a near-zone boundary 
condition, which was devised by Thorne \cite{tc69} and recently developed for 
relativistic pulsations by Lindblom et al. \cite{lmi97}, instead of the 
usually 
imposed boundary condition for asymptotically flat spacetime. 
Due to the boundary condition, complex frequencies whose imaginary 
part represents secular instability are obtained for $r$-mode oscillations 
in some polytropic models. It is found that such discrete $r$-mode solutions 
can be obtained only for some restricted polytropic models. 
Basic properties of mode solutions of equation (\ref{b-eq}) 
that obtained in this study is similar to those with the boundary condition 
for asymptotically flat spacetime although its frequency becomes complex 
because of the near-zone boundary condition.

As suggested by Lockitch et al. \cite{laf01} (see, also Refs 
\cite{bk99,y01,rk01}), when an eigenfrequency becomes 
complex number, which expresses the damping of the oscillation due to 
the energy dissipation such as the gravitational radiation emission, 
there is a possibility that the existence of the continuous part of 
spectrum in the eigenfrequency of equation (\ref{b-eq}) is avoided.
In this study, we consider the complex frequency corresponding to 
the quasi-normal mode as mode solutions of equation (\ref{b-eq}) by imposing 
the lowest order near-zone boundary condition. Our numerical 
results suggest the existence of a continuous part of spectrum in Kojima's 
equation even when the frequency is allowed to be a complex number.  
However, we still think that the existence of a continuous part of 
spectrum in axial oscillations restored by Coriolis force is not 
plausible because such property does not appear in Newtonian $r$-modes. 
The existence of a continuous part of spectrum in this study might be 
artifact due to the approximation because our treatment is the lowest 
order approximation. In other words, inclusion of full effect of 
gravitational radiation emission might avoid the existence of a continuous 
part of spectrum.  Another possibility to prevent the appearance of 
continuous spectrum might be to avoid a singular reduction in the order 
of the equation. As Kojima and Hosonuma \cite{kh00} showed, basic equations 
for $r$-mode  oscillations become a fourth order ordinary differential 
equation for metric perturbation $h_{0,l}$ when rotational effect up to 
the third order of $\Omega/(M/R^3)^{1/2}$ is consistently considered. In 
this equation, extra two degrees of freedom of solutions may be used to 
avoid singular behavior of eigenfunction. Due to these extra boundary 
conditions, all eigenfrequency may become discrete. Verification of those 
possibilities  remains future studies.

%%%%%%%%%%%%%%%%%%%%%%%%%%%%%%%%%%%%%%%%%%%%%%%%%%%%%%%%%%%%%%
%%         END OF MAIN BODY OF PAPER                        %%
%%%%%%%%%%%%%%%%%%%%%%%%%%%%%%%%%%%%%%%%%%%%%%%%%%%%%%%%%%%%%%

\acknowledgements

The authors would like to thank Prof. B. F. Schutz for his hospitality 
at the Albert Einstein Institute, where a part of this work was done. 
S.Y. is grateful to Prof. Y. Kojima and Mr. M. Hosonuma for useful 
discussion. He also would like to thank Dr. U. Lee and Mr. Y. Itoh for 
valuable comments. S.Y. acknowledges a support by a Research Fellowship 
of the Japan Society for the Promotion of Science.

%%%%%%%%%%%%%%%%%%%%%%%%%%%%%%%%%%%%%%%%%%%%%%%%%%%%%%%%%%%%%%
%%  BIBLIOGRAPHY                                            %%
%%%%%%%%%%%%%%%%%%%%%%%%%%%%%%%%%%%%%%%%%%%%%%%%%%%%%%%%%%%%%%

%

%%%%%%%%%%%%%%%%%%%%%%%%%%%%%%%%%%%%%%%%%%%%%%%%%%%%%%%%%%%%%%
%% FIGURES                                                  %%
%%%%%%%%%%%%%%%%%%%%%%%%%%%%%%%%%%%%%%%%%%%%%%%%%%%%%%%%%%%%%%

%\newpage

\begin{figure}
\epsfxsize=16truecm
\begin{center}
\epsffile{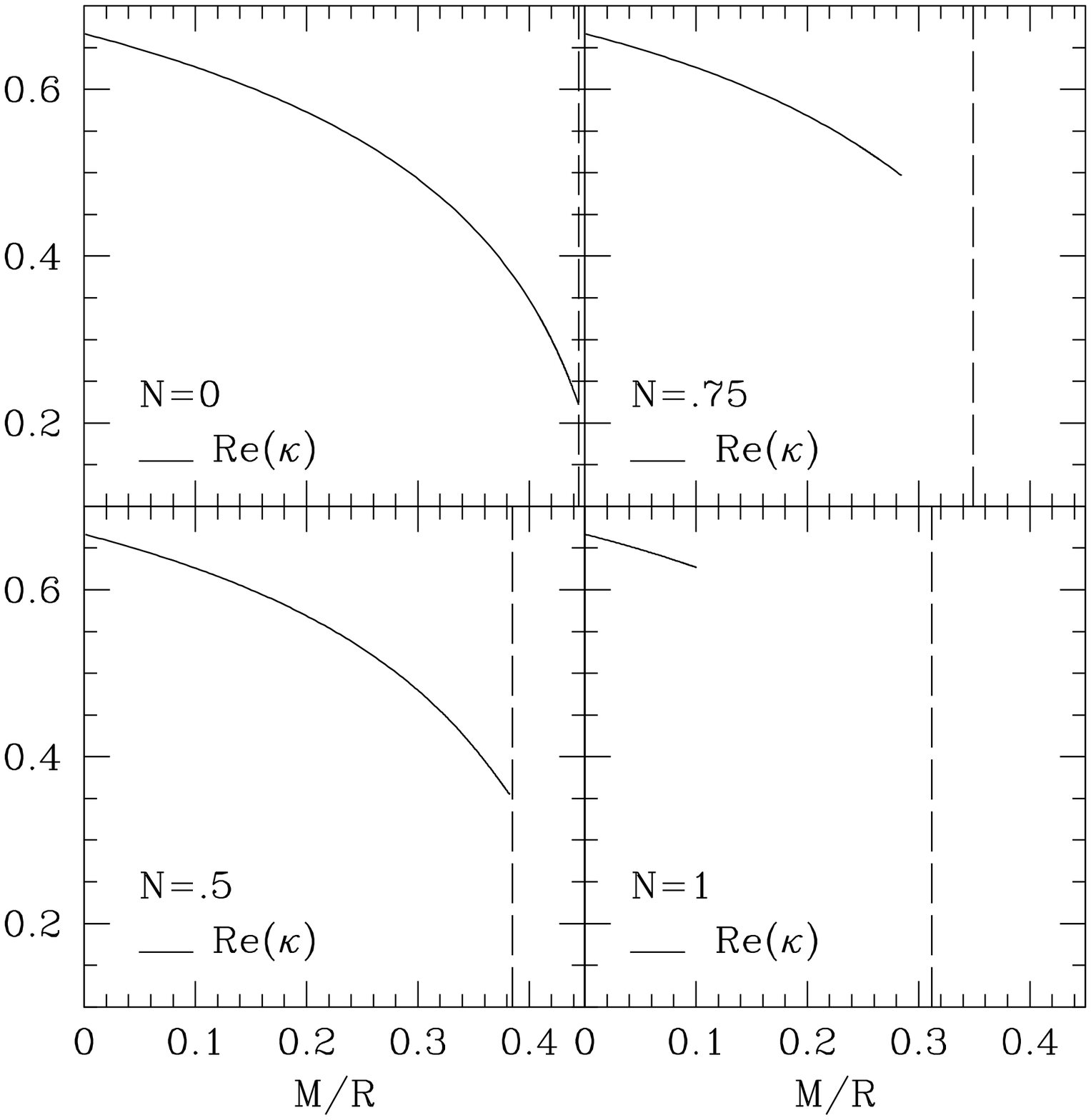}
\end{center}
\caption{Real parts of scaled frequencies 
$\kappa=\bar{\sigma}/\Omega$ of the $r$-modes with $l=m=2$ plotted as 
functions of $M/R$. In each panel, the frequencies of modes for  
polytropic models with $N=0$, $0.5$, $0.75$, and $1$ are respectively 
shown. The labels indicating their polytropic indices $N$ are 
attached in corresponding panels. Vertical dotted lines show the 
maximum values of $M/R$ for equilibrium states: $M/R=0.444$ for $N=0$, 
$M/R=0.385$ for $N=0.5$, $M/R=0.349$ for $N=0.75$, and $M/R=0.312$ for 
$N=1.0$.}
\end{figure}

\begin{figure}
\epsfxsize=16truecm
\begin{center}
\epsffile{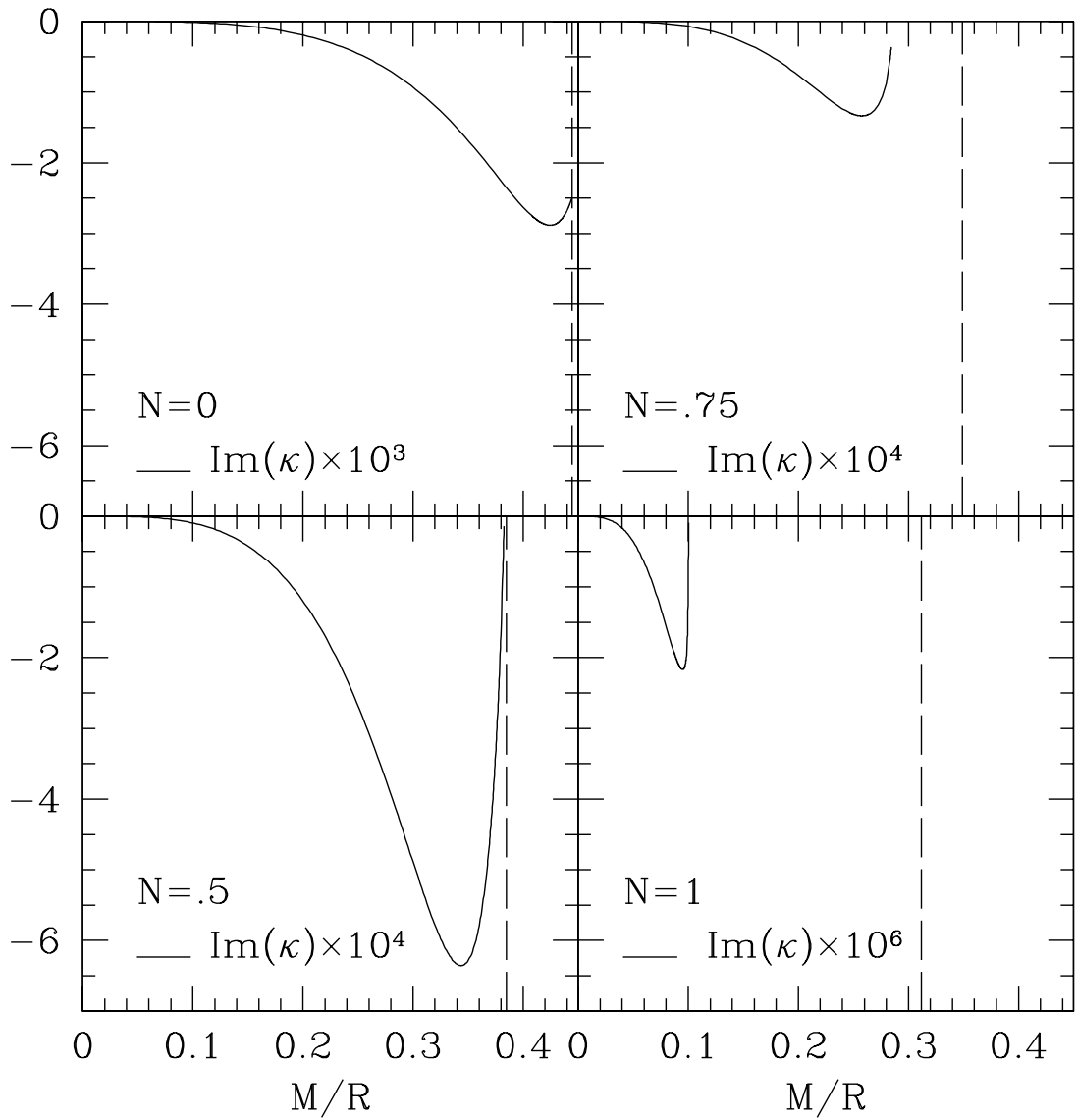}
\end{center}
\caption{The same as Figure 1 but for imaginary parts of frequency.}  
\end{figure}

\begin{figure}
\epsfxsize=8truecm
\begin{center}
\epsffile{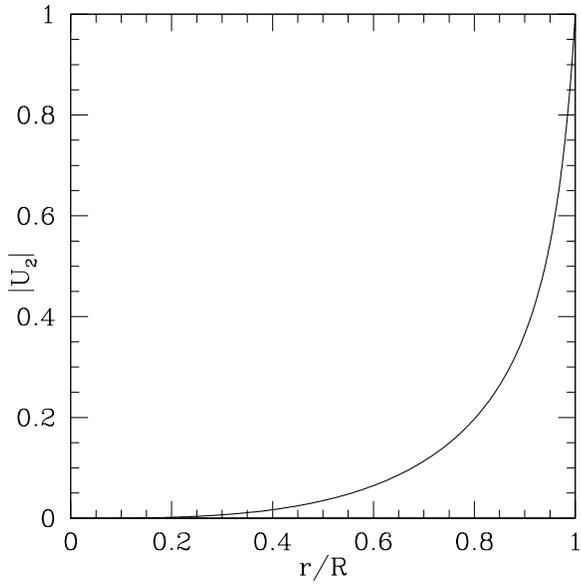}
\end{center}
\caption{Absolute value of eigenfunction $U_2$ of a $r$-mode 
with $l=m=2$ for a $N=0.5$ polytrope with $M/R=0.1$ is given as a 
function of $r/R$, where normalization of the eigenfunction is given 
as $U_2(R)=1$.}  
\end{figure}

\begin{figure}
\epsfxsize=8truecm
\begin{center}
\epsffile{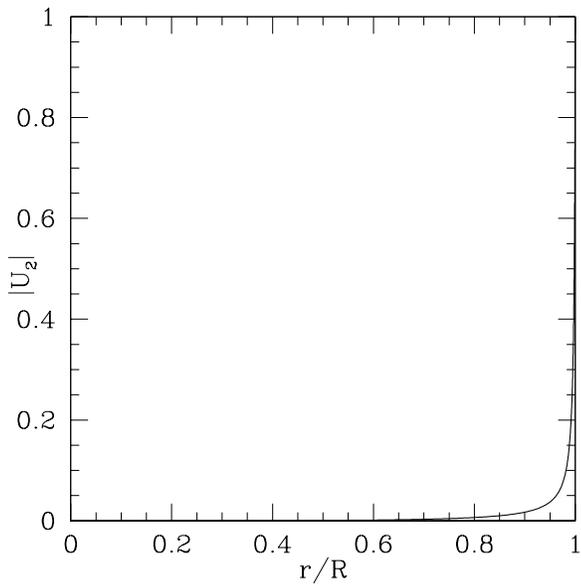}
\end{center}
\caption{The same as Figure 3 but for $M/R=0.37$.}
\end{figure}

\end{document}